# Structural biology meets data science: Does anything change?

Cameron Mura[1,3], Eli J. Draizen[1,3], and Philip E. Bourne[1,2]*

## Affiliations & correspondence

[1] Department of Biomedical Engineering; University of Virginia; Charlottesville, VA 22908; USA

[2] Data Science Institute; University of Virginia; Charlottesville, VA 22904; USA

[3] CM and EJD contributed equally to this work.

*Corresponding author: Bourne, Philip E (peb6a@virginia.edu)

## Highlights & graphical abstract

- Data science has emerged as a fourth paradigm of science, alongside the theoretical, experimental, and computational.
- Structural biology's rich history includes practices, such as an emphasis on openness and reproducibility, which can serve as positive models for many nascent areas of data science.
- Machine learning is profoundly impacting the biosciences, based on recent literature trends; we are likely at the cusp of a gold rush moment in structural biology.

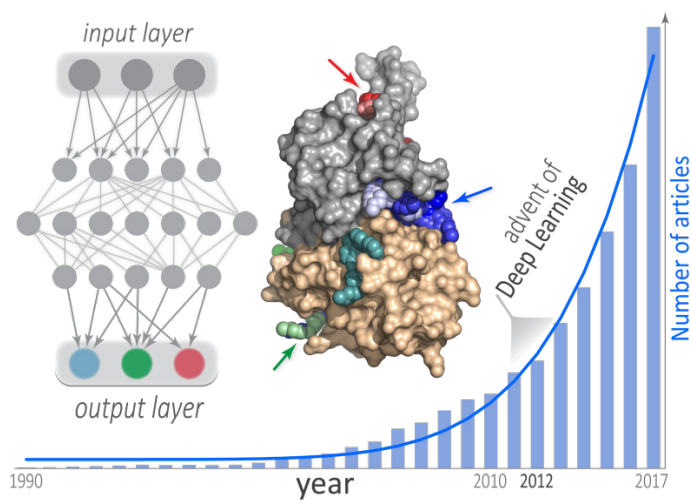

## Document information

| | |
|---|---|
| Last modified: | 22 July 2018 [revised 29 August 2018] |
| Running title: | *Structural Biology Meets Data Science* |
| Keywords: | data science; machine learning; protein interactions, topic modeling |
| Abbreviations: | CNN, convolutional neural network; cryo-EM cryo–electron microscopy; DL, deep learning; DS, data sciences; HPC, high-performance computing; ML, machine learning; NLP, natural language processing; NN, neural network; PDB, Protein Data Bank; PLI, protein•ligand interaction; PPI, protein•protein interaction; SW, software; TM, topic modelling; WMS, workflow management system. |
| Additional notes: | The main text is accompanied by 2 figures and 1 item of Supplementary Material. |
| Journal info: | *Current Opinion in Structural Biology*; October 2018; Biophysical and Computational Methods |

**Abstract**

Data science has emerged from the proliferation of digital data, coupled with advances in algorithms, software and hardware (e.g., GPU computing). Innovations in structural biology have been driven by similar factors, spurring us to ask: can these two fields impact one another in deep and hitherto unforeseen ways? We posit that the answer is yes. New biological knowledge lies in the relationships between sequence, structure, function and disease, all of which play out on the stage of evolution, and data science enables us to elucidate these relationships at scale. Here, we consider the above question from the five key pillars of data science: acquisition, engineering, analytics, visualization and policy, with an emphasis on machine learning as the premier analytics approach.

**Introduction**

The term *Structural Biology* (SB) can be defined rather precisely as a scientific field, but *Data Science* (DS) is more enigmatic, at least currently. The intrinsic difference is two-fold. First, DS is a young field, so its precise *meaning*—based on what we practice and how we educate its practitioners—has had less time than SB [1,2] to coalesce into a consensus definition. Second, and more fundamental, DS is interdisciplinary to an extreme; indeed, DS is not so much a field in itself as it is a way of *doing* science, given large amounts of diverse and complex data, suitable algorithms and sufficient computing resources. Such is the breadth and depth of DS that it has been described as a fourth paradigm of science, alongside the theoretical, experimental and computational [3,4]. Because it is so vast and sprawling, a helpful organizational scheme is to consider four *V*'s and five *P*'s that characterize data and DS (Figure 1).

     The four *V*'s describe the properties of data: *volume*, *velocity*, *variety* and *veracity*. The *P*'s are the five disciplinary pillars (P-i through P-v) of DS (Figure 1): (i) *data acquisition*, (ii) *data reduction, integration and engineering*, (iii) *data analysis* (often via machine learning), (iv) *data visualization, provenance and dissemination*, and (v) *ethical, legal, social and policy-related matters*. The *P*'s are interrelated, as are the *V*'s. For example, the fifth pillar leans into each of the other four: a host of privacy matters surround data acquisition, aggregation can have unforeseen security concerns, analytics algorithms can introduce unintended bias, and dissemination policies raise licensing and intellectual property issues. Similarly, many modes of data analysis (P-iii) rely on advanced visualization approaches (P-iv). The *P*'s also closely link to the four *V*'s. For example, P-i, the *data acquisition* pillar, clearly relates to *volume* and *velocity*. More subtle linkages also exist, e.g., between *data analysis* and *variety*: in structural biology, hybrid approaches [5-8] involve joint integration/analysis of heterogeneous varieties of data (e.g., cryo-EM, mass spectrometry, cross-linking), for instance via a Bayesian statistical formulation of the structure determination process [9,10]. The philosophy and epistemology of DS is an entire field unto itself, and helpful starting points can be found in recent texts [11].

     The rest of this review focuses on the junction of data science and structural biology. We consider DS approaches that have been applied in SB recently, including examples from crystallography and protein interactions. We focus mostly on pillar P-iii (Figure 1), and specifically machine learning. In

so doing, we largely ignore traditional disciplinary labels.  For example, the junction of DS and SB could be viewed as simply expanding the field of structural bioinformatics [12]; but, such disciplinary labels and boundaries matter less than the actual scientific impact.  Analogously, definitions of "*the internet*" vary greatly, yet the impact of the internet on science is unmistakable.  For convenience, we use the term 'SB' as including structural bioinformatics, simply to distinguish what has gone before versus what may lie on the horizon.  We suspect much lies on the horizon: akin to the rapid growth [13] of databases such as the Protein Data Bank (PDB; [14]), our assessment of bibliometric data (Figure 2) suggests that data science will profoundly impact the biosciences, including structural biology. (The best-fit curve in Figure 2 is supra-exponential, with no inflection point in sight.)  Conversely, can SB impact the broader field of DS? This has yet to occur in a definitive way, but, given the maturity of SB as a discipline, much can be learnt from it and its history; thus, we start with a short review of how SB might influence DS.

**What structural biology has to offer data science**

*Open science*

SB has pioneered open science through the provision of the PDB and many derivative data sources.  The complete corpus of structural information in the PDB is free of copyright and is available for unfettered use, non-commercial or otherwise (P-v).  Moreover, community practices—such as virtually no journal publishing an article without its data deposited in the PDB [15]—is  a precedent that, if broadly adopted in other disciplines, would deepen the amount and diversity of data available for DS-like approaches in those other scientific and technical domains.  The creation and free distribution of software (SW) tools has echoed this trend, as epitomized by the *Collaborative Computational Project 4* (*CCP4*); developed and meticulously maintained since 1979 [16], the *CCP4* suite has been a mainstay of the crystallographic structure-determination process.  *CCP4* and kindred projects, alongside myriad other SW tools and attendant data, have fostered an open discipline.  DS draws upon data and ideas from a wide range of disciplinary areas, but some of these areas have been less open than SB, at least historically.  To succeed, we believe that any DS must abide by the 'FAIR' principles, enabling researchers to *Find*, *Access*, *Interoperate* and *Reuse* data and analytics [17].  SB has exercised this for decades, and is thus positioned to lead the way.

*Reproducibility*

In principle, reproducibility is the bedrock of the scientific enterprise.  And, as a byproduct of open science, reproducibility has been central in SB, though often less so in other realms of DS.  Cultural differences across various disciplines, often driven by (perceived) competitive pressures, have dampened what could be the norm.  In SB, the systematic, pipelined nature of many structure-determination approaches has facilitated reproducibility.  A notable example is the effort, spurred by structural genomics, to annotate large-scale macromolecular crystallization experiments and to conduct careful target tracking [18]; in principle, such efforts afford a rich source of data, exploitable by DS via data mining and machine learning methods [19].

*Workflows, high-performance computing*

Reproducibility, in turn, is facilitated by workflows. Some workflow management systems (WMS) are domain-specific (e.g., Galaxy for genomics [20,21]), while others are more generic or monolithic (e.g., KNIME [22]); lightweight toolkits also exist, providing libraries to write custom parallel-processing pipelines (e.g., [23,24]). Again, structural genomics and other data-rich areas (e.g., large-scale biomolecular simulations) have prompted the development of WMS solutions. Closely related to workflows, recent technologies that have become best practices in DS—such as Jupyter notebooks (as a user interface) and Docker 'containers' (for virtualized runtime environments)—likely will be adopted more broadly in SB, as research questions become more quantitative and as data-intensive computational steps are pursued via distributed computing and other modes of HPC. Cloud computing and related approaches, such as the MapReduce paradigm (implemented in Hadoop), rapidly entered genomics and bioinformatics early on [25] and are becoming more widely adopted in other biosciences too, including SB [26]; other examples include large-scale biomolecular modeling for virtual screening and drug design [27,28] and, more recently, cryo-EM pipelines for structure determination [29].

Structural biology has relied upon HPC since the dawn of supercomputing in the 1960s. A recent example using HPC involves the phasing of diffraction data. Recognizing the wealth of structural information in the PDB, and that molecular replacement (MR) can be treated as embarrassingly parallel across all these structures, the *BALBES* [30] pipeline leverages all known 3D structures to create and then use MR search models in an automated manner. This approach was recently extended to fitting 3D models into cryo-EM maps [31]. Somewhat similar in spirit, *PDB_REDO* endeavors to automatically improve all PDB structures by re-refining 3D models against the original X-ray data, utilizing established refinement approaches (e.g., TLS) and grid computing [32,33]. As a final example, a recent and highly creative approach to crystallographic phasing has a strong DS feel: Encoding phase values as 9-bit strings (genes), and applying a genetic algorithm for sampling/optimization, Yeates & colleagues [34] developed a crowdsourced gaming platform for *ab initio* phasing, at least to low resolution. Such 'citizen-science' [35,36] approaches will likely play broader roles in SB (and DS) in the coming decade.

*Visualization*

Visualization has played a key—indeed, *defining*—role in SB since the 1950s, when the first macromolecular structures were determined. Concepts, principles and best practices for biomolecular visualization can be found in many reviews [37-39]; the supplement in [37] traces the historical development of this field. Recent advances have occurred in web browser-embedded, hardware-accelerated tools for interactive molecular visualization, such as the NGL Viewer [40]; in the future, a greater share of visualization work likely will occur within browsers. To transcend how molecular renderings are usually communicated (as static images), we suspect that much could be gained by comparing visualization techniques in DS and SB. Though iconic and highly informative, beware the "curse of the ribbon": macromolecules are dynamic, multifaceted entities, and static renditions are but a starting point. For similar reasons, there is a need for molecular visualization platforms that transcend simple graphical viewers—that enable facile, flexible and extensible integration of other forms/modalities of data and novel visualization techniques [41], such as the data-intensive sort that often arise with big data. We believe that DS tools can address this need; note that texts are becoming available on this topic, such as the recent *Big Data Visualization* [42]. Ideas and methods from beyond SB—such as "chord diagram" layouts in genomics [43], termed "hierarchical edge bundles" [44] in computer graphics—can be applied in SB, for instance to visualize data associated with hierarchical clustering of protein structural differences (e.g., see the figures in [45]).

Finally, note that further areas of SB×DS overlap can be identified, but are not treated here because of space limitations. Three such examples are: (i) database (DB)-related issues, including structured versus unstructured data, relational versus non-relational DBs and query languages [46]; (ii) systems and network biology [47]; and (iii) ontologies and formal knowledge representation systems [48,49].

**What data science analytics has to offer structural biology**

DS analytics spans a vast territory, including applied mathematics, statistics and computer science. Here, we focus on two machine learning (ML) approaches—one which has received much recent attention (deep learning [DL]), and one for which we envision possible applications in SB (natural language processing [NLP]). A glossary is included (Box 1), and the Supplementary Material offers: (i) a brief primer on ML, (ii) a concise historical note on early applications of neural networks in SB, and (iii) a short sketch of the general applicability of data sciences in structural biology and other biosciences.

*Machine learning applied to biomolecular interactions*

In a recent wave of activity, DL methods have been applied to model and predict protein•ligand and protein•protein interactions (PLI, PPI). Deep learning is a type of ML that employs deep (multi-layered) neural network (NN) architectures; training and deploying such architectures is now feasible because of the exceptional computational performance of modern GPU-equipped clusters.

Accurately predicting and modeling PLIs (structural poses, energetics) would advance many areas, both basic (e.g., evolutionary analyses of ligand-binding properties) and applied (e.g., drug design and discovery). Historically, this field has largely relied on two distinct methodological approaches: quantitative structure-activity relationships (QSAR) and *in silico* docking. Virtual screening, wherein one docks against large libraries of small compounds, is an established example of DS in SB; as a newer example, note that workflow-based approaches to high-throughput crystallographic fragment screening have a significant DS component [50]. Extensions of the basic QSAR and ligand-docking approaches also call upon DS. For example, recognizing that a protein exists as an ensemble of thermally-accessible conformational states in solution, simulations have been combined with docking in the "relaxed complex" scheme to capture receptor flexibility [51]. Similar in spirit, data-intensive "ensemble-based" methods [52] can enable dynamic pharmacophore models (e.g., [53]) to be devised. In a recent approach, a workflow to discover 'cryptic' (and druggable?) binding sites was developed by integrating comparative structural analyses, pocket-detection algorithms, fragment docking, molecular simulations, and an ML classifier [54]. In another data-driven, structure-based approach, Zhao et al. [55] recently analyzed the human kinome by integrating ligand-binding data with protein-ligand "interaction fingerprints" and a sequence order-independent profile–profile alignment method ([56]; useful for determining specificity among similar ligand-binding sites).

Recent work on predicting PLIs has directly employed ML, including for the interrelated goals of *virtual screening*, *affinity prediction* and *pose prediction*. The application of statistical and ML approaches, in particular deep neural nets, to the PLI problem was reviewed recently [57]. Here, we mention only that a surge of new work has applied convolutional neural nets (CNNs) to the PLI problem—references [58-63] comprise a partial list from just the past year. Notably, these purely ML-based approaches rely on human expertise only in the early stage of choosing structural descriptors (hydrophobicity, ionizability, etc.), which are input features for NN training. The protein structure (either as a complex, or just receptor) can be treated as a 3D image, wherein atoms that compose the structure are assigned to discrete volumetric elements (voxels). CNNs excel at learning from 2D image data [64], suggesting that their 3D counterpart, 3D deep CNNs, can be used for volumetric analysis. Leveraging these ideas, the 3D DCNN of *DeepSite* achieved state of the art performance, having been trained on known protein•ligand structures [58].

As with PLIs, protein•protein interactions (PPI) are critical to much of cell biology, and are another focal point of recent ML efforts. Were all binary PPIs known, they could be used to build whole species interactomes [65] and inter-species (e.g., host-pathogen) interactomes [66,67], which, in turn, would aid elucidation of signaling pathways [68], metabolic networks (Recon3D [69]), and evolutionary pathways [70]. ML can be used to predict which two proteins interact and what specific residues ('hot-spots') mediate the interaction (i.e., binding sites). If both binding sites (or interfaces) are known, they can be used to model structures of their complexes.

Thus far, the optimal information for predicting interacting residues has been at the sequence level, using residue co-evolution. Intuitively, residues that co-evolve between two proteins are likely to contact one another. Such sites can be predicted using ML and DL methods like maximum entropy models or 2D-CNNs [71-75]; a drawback to such approaches is the need for sufficiently large protein families. For purposes of structure prediction, the same approach can be used to predict residue•residue contacts from one protein family alignment. One can also predict PPIs from structure if a query protein is homologous (based either on sequence or structure) to one protein in a known PPI. If the identity of only one interaction partner is known, and the binding sites in the other partner unknown, binding sites and partners can be predicted by structurally aligning a query to crystal structures of complexes, using either local (e.g., PRISM [76]) or global (e.g., IBIS [77]) 3D superimpositions. Residues from the query protein that align to one side of an interface are predicted to be a part of the binding site.

ML methods can also predict binding-site residues given the 3D structure of only one partner. Here, atomic and residue-level features (e.g., hydrophobicity, phylogenetic conservation) are calculated for all structures in the PDB. True binding site residues are taken from crystallized complexes, split into monomers, and used to train a classifier (SVMs, Decision Trees, etc.). Unfortunately, such predictors have suffered from low precision and recall [78], perhaps because the 3D spatial details of the data are not retained but rather enter the model only as 'flattened' features (or, assumptions of independence are applied incorrectly). Given current limitations and difficulties, it is unsurprising that DL is now starting to be applied to PPI modeling and prediction. While only biomolecular interactions are discussed here, we envision that contemporary DL approaches, such as variational autoencoders, will play major roles in areas such as structure prediction [79], protein design and evolutionary analyses [80].

*Natural language processing applied to biomolecular assemblies*

NLP is a form of machine learning concerned with processing and analyzing language, written or spoken. Here, 'processing' can mean many things: analyzing frequencies and co-occurrences of words and higher-order units (*n*-grams), parsing texts in syntactic/grammatical analyses, information retrieval, machine translation, language *comprehension* (and *synthesis*), and beyond. The scope of this topic is vast, and helpful biosciences primers are available [81-83]. NLP methods play key roles in routine tasks such as *search/query* (databases, knowledge-bases), *information extraction* and *text summarization*; indeed, you may have reached this article via a web-query using the PubMed search engine. PubMed is but one example of NLP in biomedical informatics, and it remains an active area of research; recent examples include a "neural word embedding" approach for document matching in PubMed [84] and development of a flexible *term* ↔ *concept* matching system for biomedical pipelines [85]. Apart from search and mining in biomedical literature, might NLP impact structural biology in other ways?

NLP-like approaches have been applied to detect the subcellular localization of proteins [86,87] and, recently, to predict structures of protein complexes [88]. Notably, using ML–enhanced NLP, versus a purely text-mining–based NLP approach, was found to significantly improve the structural predictions of complexes [89]. Note that both sorts of problems—subcellular localization and structural modeling—are distinctly spatial, or image-based, as opposed to textual. For this reason, we expect that a relatively new and highly-generalized approach to NLP, termed topic modeling (TM), holds great promise in the biosciences. In TM, 'topics' are extracted over a corpus of unstructured data (e.g., a set of books) using a probabilistic machine learning framework; fundamentally, this is achieved by examining the distributions of words (a "bag of words" ansatz) under a generative statistical model, such as the latent Dirichlet allocation (LDA). An introductory review of TM and a recent overview of TM-like approaches in bioinformatics can be found in refs [90] and [91], respectively. To extend TM to other areas—including even the learning of topics (themes) from non-textual data like protein structures—the basic issue is one of defining a suitable mapping of one's problem to TM's core framework of *document* ↔ *topic* ↔ *word*. As a potential horizon, we suggest that TM may be applicable to the analysis of protein folds and other biomolecular structures. Such an application of NLP to what is a fundamentally geometric problem finds precedent in the pioneering development of a generative Bayesian hierarchical model for scene classification from raw image data [92].

**Conclusion**

In addressing the topic posed here—SB meets DS—we have considered the influences of these fields on one another, given their respective stages of maturity. SB's rich history could positively influence the five pillars of DS (Figure 1). For example, data collection and processing often entails well-established workflows, standards and practices (e.g., structure validation), such that the results (3D structures) can be taken as "gold standard" data in downstream analyses. Moreover, SB uses some approaches, like ontologies (for standardization, automated relationship discovery), that are not as prevalent in DS, but which could enjoy broader application. Notably, the data-access and software-sharing policies that have evolved in SB communities for decades can serve as positive models for DS.

Conversely, DS is being driven by economic, political and social factors that reach far beyond science itself (technology, commerce, etc.), and which inspire scientists to further innovate across the five pillars of DS. The links to SB are many and varied, and here we have touched on but two of them (ML more broadly, and NLP in particular). Details of ML and NLP approaches are beyond the scope of this work, which has only sought to briefly sketch potential synergies between these DS-based approaches and SB.

We answer our question, then, with a resounding yes: DS is already impacting SB, and we argue that the converse could be true, too. The rate of change is less certain, but is clearly steep: Figure 2, based on the recent biosciences literature, suggests that we are at the cusp of a major impact. Realizing the full benefits of this gold-rush moment will require more multidisciplinary training of students, support from scientific bodies, deep funding and, most importantly, a general willingness by the respective scientific communities. These are interesting times indeed.

ACKNOWLEDGEMENTS

We thank D Mietchen, S Veretnik and Z Zhao (UVa) for helpful discussions and critical reading of the manuscript. Portions of this work were supported by the University of Virginia (PEB) and National Science Foundation CAREER award MCB-1350957 (CM).

CONFLICT OF INTEREST STATEMENT

The authors declare no conflict of interest.

**Glossary**

The following terms, organized here thematically, appear in this review or are pervasive in the literature. As part of the DS jargon, the terminology may be unfamiliar and is therefore included for convenience.

Statistical and machine learning

- *Classifier*: An algorithm or function that maps input data into one of at least two categories (or *classes*). For instance, if only two classes are possible (e.g., *True* or *False*, *Even* or *Odd*), and our input data are integers, then the *modulo* operation (*mod* 2) could serve as a *binary classifier*.

- *Model*: A formal relationship between input data and some set of outputs; another way to view this is as a *mapping*, *association rule* or mathematical *function*. As a concrete example, say we have an ideal, one-dimensional spring on a frictionless surface. Say we collect dense (finely-sampled) data on the precise position ($x$) of the spring's terminus at many time-points (i.e., we have a time-series, $\{x(t)\}$). To elucidate the system's behavior in terms of our data, we may propose an equation, say the sinusoid $x(t) = A \cdot \cos(\omega_0 t + \phi)$, where $A$ is amplitude, $\phi$ is phase and $\omega_0$ is angular frequency. This functional form is what we mean by a *model*: the precise parameters will vary from system to system (different springs, stiffnesses, etc.), and what matters instead is the *functional form* of the mapping (in this case, the equation of motion *models* simple harmonic oscillation). For more complex systems—e.g., recognizing patterns in images, delineating protein structures—such simple, closed-form expressions generally do not exist (nevermind us being able to propose them *a priori*!); statistical approaches come to the rescue by offering a way to learn a model for the set of *input* ↔ *output* associations.

- *Regression*: A statistical approach to estimate relationships amongst variables; e.g., linear regression will estimate a linear relationship (or slope) between two or more variables, which can be used for purposes of prediction and classification.

- *Random forest*: An ensemble of *decision trees*.

- *Decision trees*: A classifier that follows *if–then–else* decision rules to traverse a directed graph, thus predicting an output. The rules, or nodes of the tree, are the features of the model sorted by information gain when split on certain values of the features.

- *Supervised, unsupervised learning*: See the Supplementary Material for a description of these terms.

- *Support vector machine* (SVM): A classifier that finds a linear discriminatory boundary between classes, generally via regression in a higher-dimensional space or application of kernel methods (most simply, a '*kernel*' can be viewed as a measure of similarity between two feature sets, e.g., the dot product).

Neural networks and deep learning

- *Neural network (NN)*: Also known as an 'artificial NN' or 'multilayer perceptron' in the older literature, these are mathematical networks of nodes, which are the processing units (loosely, neurons; also termed 'hidden units'), and edges, which link the nodes. All NNs consist of at least two layers that interface with the environment: an *input layer* of nodes (receives input data) and an *output layer* (emits processed data [i.e., predictions, results]).

- *Feedforward NN*: A NN architecture wherein information flows through the network unidirectionally, from the input layer to the output layer. This is possible because the edges (links) are directed from one node to another; this network topology is a type of directed acyclic graph (DAG), and other DAG-based NN architectures are conceivable.

- *Convolutional NN (CNN)*: A NN that applies convolutional operations, which take local, connected, sub-regions of an input matrix as neurons. Inputs are typically 2D images, which is a 2D matrix of pixels, where the sub-regions are smaller pieces of the image, or 3D volumes where smaller cubes traverse the volume.

- *Deep NN (DNN)*: Most simply, a NN architecture that includes multiple hidden layers.
- *Backpropagation*: A method to update learnable weights of the NN interconnects between nodes by transmitting errors backwards (in the direction from the output layer towards the input layer); this backwards propagation of errors, in turn, corresponds to the network improving as a predictor, i.e., the network can be said to 'learn'. More concretely, backpropagation proceeds by applying the chain rule to compute the gradient of the error (the *loss function*) at each filter (node) for a given layer, and iteratively using the gradient values to update the weights; therefore, this is fundamentally a gradient descent algorithm, as found in many classes of optimization problems.
- *Loss function*: A function to compute the error between the true and predicted values. For example, this could be as simple as the Euclidean distance between estimated and true (target) values.
- *Dropout*: A technique to address overfitting by removing a randomly selected subset of nodes, in a single layer, during training (a forward and backward pass). This allows the NN to learn more robust features by testing different possible subsets of nodes; typically, on the order of 50% of nodes are silenced.
- *Regularization*: A technique to optimally balance the perils of underfitting/overfitting to training datasets.
- *Epoch*: One forward and one backward pass of all training data. Many epochs (typically ranging from 30-1000) are usually required before a NN model converges.

## Natural language processing

- *Corpus*: Most simply, a collection of information. This term, prevalent in the NLP field, is frequently used to generically refer to written data (books, journals, etc.), where it often means a comprehensive collection on a particular topic (all writings by particular authors, or about a particular protein, etc.).
- *Topic*: Most simply, a statistical distribution of words, each word being drawn from a well-defined set of words (a *fixed vocabulary*); a *topic* can also be viewed as a *theme*. In many ways, a given document is defined by its collection of most prevalent topics.
- *Topic modeling* (TM): A set of unsupervised algorithms to discover the topics in a corpus of (unstructured) information, generally by applying statistical algorithms to analyze and model word distributions.
- *Latent structure*: A highly general concept, referring to there being some general correlation (or nonuniformity, or 'structure') among the hidden (latent) random variables that define the probabilistic distributions underlying models such as LDA. In TM, we seek to learn these relationships between hidden variables (i.e., the *structure*), which manifests in the form of (non-random) *topics*.
- *Latent Dirichlet allocation* (LDA): A type of TM wherein a document (a "bag of words") is viewed as a probabilistic distribution over a set of topics; a topic, recall, is a distribution over words. A rather involved generative statistical model underlies LDA; a 'generative' model means that the observed data (the document, its words, their distributions) are taken as having been *generated* via sampling a hidden distribution (a random process, or, if there is latent structure, a non-random process). Briefly, each document's set of topics are taken to be *Dirichlet–distributed*, and the words in a document are *allocated* to its various topics based on this distribution. (In the Bayesian sense of *joint distributions*, *priors*, etc., the Dirichlet distribution is the *conjugate prior* to the multinomial distribution that is taken as explaining the distribution of topics.)

**Figure Captions**

Figure 1. SB mapped onto the five pillars of DS, and in relationship to the four V's of big data. DS rests upon five central pillars, schematized in (a) as (i) data acquisition; (ii) data integration & engineering; (iii) data analytics (e.g., machine learning); (iv) visualization, provenance and dissemination; and the (v) ethical, societal, legal and policy aspects. General concepts and keywords from the data sciences are near the bottom of each column (e.g., MapReduce, a distributed computing paradigm), while more domain-specific examples rest atop each column (e.g., structure–based drug design [SBDD], middle column). A band of opportunity arises as SB meets the data sciences. Realizing these potential opportunities requires big data, which enables a question or system to be addressed via DS approaches like deep learning. The four V's of big data—*volume*, *velocity*, *variety* and *veracity*—are shown in (b), illustrated by vignettes from SB. As indicated, the *volume* and *velocity* characteristics are intertwined; for instance, modern X-ray diffraction technologies enable shutter-less data collection, with upwards of many millions of diffraction patterns acquired per day (a concomitant increase in the rate of structure determination means growth in the volume of the PDB). Fits of the data in the PDB histogram (b) to different functional forms—(i) a simple power law, (ii) a pure exponential, (iii) a stretched exponential and (iv) the product of an exponential and a power law—reveal form (iv) to be the best fit (orange trace). The *Variety* panel illustrates the challenge addressed by 'hybrid methods': data arise from cryo-EM, X-ray diffraction, NMR spectroscopy, molecular simulations, chemical cross-linking/mass spectrometry, phylogenetic analyses and a host of other potential approaches. DS provides a framework for integrating such data in an optimal manner (in an information theoretic sense) so as to create 3D structural models.

Figure 2. **The recent surge in publishing activity for machine learning in the biosciences**, shown here via bibliometric data obtained for the PubMed/MEDLINE (orange) and ISI Web of Science (blue) literature databases. The overlaid histograms show the number of publications in which the string "`machine learning`" co-occurs in the title or abstract fields; the precise PubMed query was "`machine learning[Title/Abstract]`", and the string "`machine learning AND bio*`" was used for an ISI Topic search. Both datasets were fit with the same four functions listed in [Figure 1](). For both PubMed and ISI, a subtle crossover occurs wherein a supra-exponential (form iv) gives a better fit than a pure power law; such highly nonlinear 'J-curves' or 'hockey stick curves' arise in systems subject to singularly disruptive forces (e.g., human population growth after the Industrial Revolution, climate temperatures in the past century). Intriguingly, the approximate year of crossover—2010 for the ISI data, 2012 for PubMed—is generally regarded as the "breakthrough year" for Deep Learning (e.g., Google Brain learned a 'cat' *de novo*, from YouTube data), enabled by advances such as GPU computing, algorithmic approaches such as ReLU and 'dropout', and vast stores of labelled training data (ImageNet). Judging by these charts, ML has begun driving a substantial transformation in the biosciences.

**References and recommended reading**

**Figure 1.** Structural biology mapped onto the five pillars of data science, and in relationship to the four *V*'s of big data.

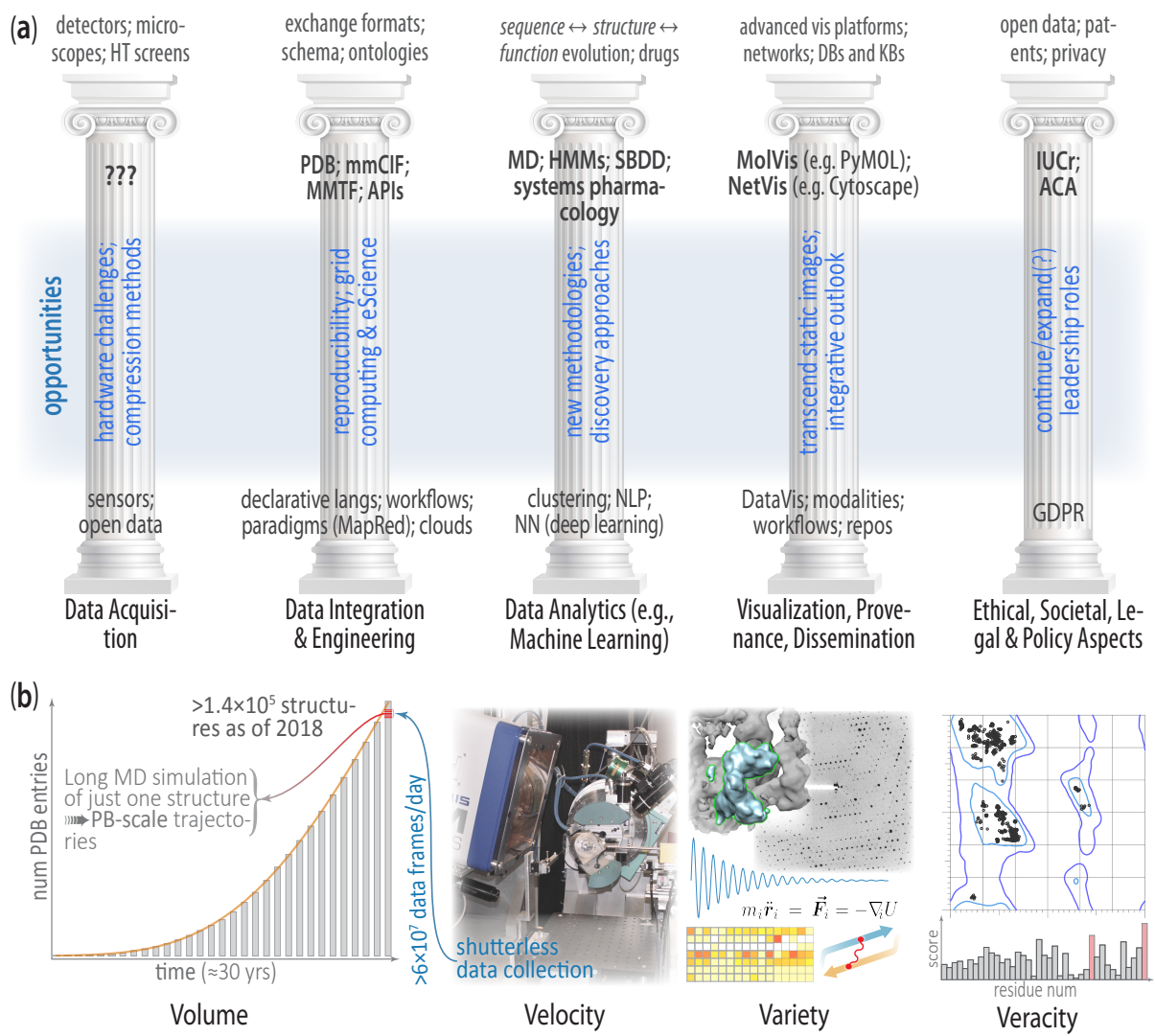

**Figure 2**. The recent surge in publishing activity for machine learning in the biosciences.

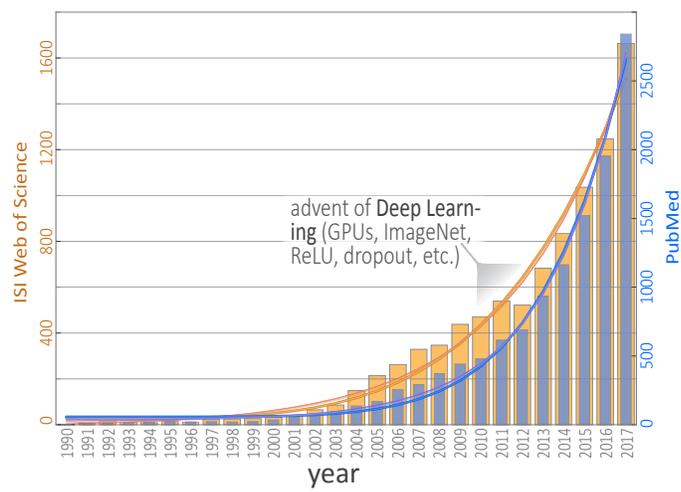

<span style="color:blue">Supplementary Material</span>

# Structural biology meets data science: Does anything change?

Cameron Mura[1,3], Eli J. Draizen[1,3], and Philip E. Bourne[1,2]*

[1]Department of Biomedical Engineering; [2]Data Science Institute;
University of Virginia; Charlottesville, VA 22904; USA
[3]CM and EJD contributed equally to this work.
*Corresponding author: peb6a@virginia.edu

This **Supplementary Material** provides (i) a brief primer on machine learning, (ii) a concise historical note on early applications of neural networks in structural biology, and (iii) a short sketch of the general applicability of machine learning/DS-based approaches in structural (and other) biosciences.

A brief primer on machine learning

Machine learning (ML) emerged from efforts in the artificial intelligence (AI) communities of the 1960s. With its possibilities and promises oversold, AI went on to largely languish in the ensuing decades. A resurgence occurred in the foundations of statistical learning theories and algorithms in the 1980-90s; coupled with advances in computing power in the past decade, this resurgence yielded a silent revolution in ML from the 1990s to the early 2000s. ML has advanced so significantly in the past decade that, today, it is often taken as synonymous with AI. Data-rich scientific disciplines, such as the biosciences (and particularly structural bioinformatics), have increasingly adopted ML approaches, driven by (i) improvements in algorithms, (i) software libraries and implementations that have become more accessible to non-specialists, (iii) training data that have become richer in complexity and more abundant, and (iv) remarkable strides in commodity computing power, chiefly via graphics processing units (GPUs) and approaches such as general-purpose computing on GPUs (GPGPU). ML enjoys great visibility because of its successes in pattern recognition, computer vision, image classification, difficult games (e.g., Go [1], which has a high branching factor), and various types of natural language processing (information retrieval, machine translation, etc.). The ability of an algorithm to 'learn' directly relates to the quality, complexity and availability of the data from which it learns. As a cautionary note, life sciences data are complex, with many potential confounders; recognizing these limitations will enhance any application of ML to structural biology (SB).

    How do ML approaches 'work' to model a system? (What it means, most generally, to 'model' a system is described in the Glossary that accompanies the main text.) First-principles, physically-grounded theories are intractable for systems as complex as those encountered in biology, and the core premise of ML is to take a wholly different approach. Rather than force models on data (e.g., a harmonic oscillator to model bond vibrations), the approach is to allow models of a system to emerge (be learned) from the

data. That is, the defining feature of ML is its focus on algorithms that can learn from (and make predictions based on) data. That is why ML is so central in data analytics.

Data, in turn, are central in ML because they can be used in advanced statistical frameworks and probabilistic algorithms to model (or learn) literally any system [2]. More concretely, to 'model' means to learn some function, $f$, that maps $f: \mathcal{X} \mapsto \mathcal{Y}$. With enough data and sufficient sampling, statistical methods can learn associations between inputs ($\mathcal{X}$) and outputs ($\mathcal{Y}$). Indeed, a sufficiently well-sampled system can be viewed as nearly synonymous with the data describing it. In addition to the basic statistical approaches to be applied, also required are: (i) large volumes of data, (ii) an objective/target function to train the ML system, sometimes referred to as a *loss*, *cost*, or *fitness* function, and (iii) an algorithm to sample the solution space, typically to find extrema of the objective function; the algorithm drives the key training/learning stage. The word 'algorithm' is used in a quite general sense in ML: it can be conceptually straightforward, as with the idea of a genetic algorithm, or it may correspond to something fuzzier, such as the directional flow of information (data, weights, etc.) in a feed-forward neural network (NN). In NNs, the network, with its weight update scheme and other parameters, *is* the algorithm. In NNs, the learning algorithm often comes from a class of iterative optimization methods; stochastic gradient descent, with backwards propagation of errors ('*backprop*') to update weights, is one such training method. Some of the terminology in this field is provided in an accompanying Glossary.

A fundamental distinction between ML algorithms is whether they are *supervised* or *unsupervised*, and a related issue is *labelled* versus *unlabelled* data. Systems typically analyzed by ML are characterized by data that populate high-dimensional, multi-parameter spaces (hence the need for big data). A supervised learning method is trained against reliable, labelled data (e.g., if an image is a 'cat', 'lion', 'dog', etc.), and then the trained model can be used to classify unseen input data. The two basic types of learning— supervised and unsupervised—fall naturally along the labeled/unlabeled divide: a learning method is said to be *supervised* if it is trained against labelled target data prior to production usage (NNs are a prime example), whereas an *unsupervised* algorithm or *classifier* 'learns' (detects) any inherent/latent structure in unlabelled input data (in addition to NNs, clustering is an example of an unsupervised approach often encountered in SB, e.g. [3]).

Early applications of neural networks in structural biology: A concise historical note

ML's historical roots in AI reflected particular types of goals: major areas of early (and ongoing) activity included pattern recognition (e.g., in speech), computer vision, image classification, and information retrieval (text mining and related fields, such as machine translation). Notably, the widely-recognized applications of neural networks and other ML approaches in those areas (see, e.g., [4] for an old review) were concurrent with many of the first forays of NNs in structural biology—early examples include the predictions of protein secondary structures [5], transmembrane helices [6], signal peptides and other sorting signals [7], and subcellular localization of proteins [8].

General applicability of ML/DS-based approaches in structural (and other) biosciences: A short sketch

The broad applicability and general efficacy of machine learning/data science–related approaches in structural biology is closely tied to one of the V's of big data—namely, *Variety*. Variety is easily understood, though not often easily addressed, in structural biology. In short, variety refers to the various forms of data being considered (generated, transmitted, aggregated and otherwise processed). A hallmark of modern, data-intensive analyses in structural biology, as well as other biosciences (and, indeed, in any scientific or technical realm), is that the data are typically of multiple disparate types, and we seek a way to leverage the intrinsic information content of each of those types in order to achieve a goal—be it a decision (in business analytics or marketing) or an improved representation or model for a system (in structural biology).  The issue of *types* or *modalities* of data is crucial—how might one handle heterogeneous (and potentially large) sets of data?  Here, 'handle' does not mean simply the act of data-wrangling (a major effort in DS, in and of itself [9]), but rather how to most effectively 'combine' or utilize the various types of data to allow one to formulate more complete, accurate and predictive models than would be otherwise possible (with only a single type of data/information)?  This, in essence, is arguably the key goal in all the various domains to which DS is applied: we want *predictive* models (actionable, and testable/verifiable/falsifiable).  Ideally, the models are *interpretable*, too, in terms of some underlying physical theory or molecular principles (that, indeed, is a gripe sometimes lodged against the 'black box' aspect of ML approaches such as neural networks).  This general topic is precisely where the variety 'V' of data science  can flourish in structural biology and more broadly in the biosciences: a central characteristic of DS approaches (like most ML approaches) is that they generally provide a data-analysis/problem-solving *framework* that is highly generalized (agnostic of the particular problem domain), that is built upon a well-principled statistical foundation (e.g., usage of maximum likelihood estimation in crystallographic phasing and refinement [10]), and that is abstracted away from the details of the particular problem at hand.  It is for this reason that, for instance, decision trees and random forests can be applied to problems as diverse as enzyme function prediction [11], structure-based prediction of protein-protein interfaces [12], and RNA splice-site recognition [13].  And, it is precisely this principle that makes data science so powerful for integrative/hybrid methods for structure determination.